\documentclass[]{iopart}
\usepackage{iopams,graphicx,hyperref}
\usepackage[usenames,dvipsnames]{color}
\usepackage{mathrsfs}
\DeclareSymbolFontAlphabet{\mathrsfs}{rsfs}

\newcommand{\tfrac}[2]{\textstyle \frac{#1}{#2}}
\newcommand{\half}{\tfrac{1}{2}}
\newcommand{\third}{\tfrac{1}{3}}
\newcommand{\fourth}{\tfrac{1}{4}}
\newcommand{\four}{{}^{(4)}}
\newcommand{\Lie}{\mathcal{L}}
\newcommand{\cC}{\mathcal{C}}
\newcommand{\cS}{\mathcal{S}}
\newcommand{\const}{\mathrm{const}}
\newcommand{\talpha}{{\tilde\alpha}}
\newcommand{\tU}{{\tilde U}}
\newcommand{\tKrth}{{\tilde K^r{}_\theta}}
\newcommand{\tW}{{\tilde W}}
\newcommand{\hW}{{\hat W}}
\newcommand{\tA}{{\tilde A}}
\newcommand{\cs}{\tfrac{c}{s}}
\newcommand{\eplus}{\rme^{s\eta}}
\newcommand{\eminus}{\rme^{-s\eta}}
\newcommand{\epluss}{\rme^{2s\eta}}
\newcommand{\eminuss}{\rme^{-2s\eta}}
\newcommand{\Scri}{$\mathrsfs{I}^+\,$}
\newcommand{\hateq}{\; \hat{=} \;}
\newcommand{\rmin}{r_\mathrm{min}}
\newcommand{\Gammaz}{\Gamma\hspace{-0.25cm}{}^{\mbox{\r{~}}}{}\hspace{-0.12cm}}

\begin{document}

\title{An axisymmetric evolution code for the Einstein equations on
  hyperboloidal slices}
\author{Oliver Rinne}
\address{Department of Applied Mathematics and Theoretical Physics,
  Centre for Mathematical Sciences, Wilberforce Road, Cambridge CB3 0WA, UK 
  \\and\\King's College, Cambridge CB2 1ST, UK
}
\ead{O.Rinne@damtp.cam.ac.uk}

\begin{abstract}
  We present the first stable dynamical numerical evolutions of the Einstein
  equations in terms of a conformally rescaled metric on hyperboloidal
  hypersurfaces extending to future null infinity.
  Axisymmetry is imposed in order to reduce the computational cost.
  The formulation is based on an earlier axisymmetric evolution scheme, 
  adapted to time slices of constant mean curvature.
  Ideas from a previous study by Moncrief and the author are applied
  in order to regularize the formally singular evolution equations at
  future null infinity.
  Long-term stable and convergent evolutions of Schwarzschild
  spacetime are obtained, including a gravitational perturbation.
  The Bondi news function is evaluated at future null infinity.
\end{abstract}

\pacs{04.20.Cv,  
      04.20.Ha,  
      04.25.D-,  
      04.30.-w   
}



\section{Introduction}
\label{s:Introduction}

A major problem in numerical relativity that is currently attracting
considerable interest is the global treatment of entire asymptotically
flat spacetimes.
This is relevant both to the modelling of the gravitational
radiation emitted by compact astrophysical sources, and to
several questions in mathematical relativity such as the nonlinear 
stability of black holes and the quantitative asymptotic behaviour of
perturbations of such spacetimes.

In the standard approach based on the Cauchy formulation, one
truncates the spatial domain at a finite distance from the source,
where boundary conditions must be supplied (see \cite{Sarbach2007} for a
review).
Apart from being compatible with the Einstein constraint equations and
yielding a well-posed initial-boundary value problem, these boundary
conditions must carefully take into account the propagation of
gravitational radiation.
For instance, crude choices of boundary conditions are known to lead 
to incorrect results on radiation tails in the evolution of
perturbed black hole spacetimes \cite{Allen2004,Dafermos2004}.
Recently, improved absorbing boundary conditions that minimize spurious
reflections of gravitational radiation have been derived and
implemented \cite{Buchman2006,Buchman2007,Rinne2008b}.
While constituting a significant improvement,
this approach still relies on the assumption that the Einstein equations may
be linearized about a given background spacetime so that incoming and
outgoing radiation can be identified---an assumption that may not
always be valid \cite{Deadman2009}.
In the full nonlinear theory there is no satisfactory quasi-local definition
of a gravitational energy flux at a finite distance from the 
source \cite{SzabadosLRR}.
This fact also complicates the extraction of gravitational
radiation in an invariant manner.

Gravitational radiation is only unambiguously defined at
future null infinity \Scri \cite{Bondi1962,Sachs1962}.
Thus it is very desirable to include \Scri in the computational domain.
There are several ways of doing this.
One approach is characteristic evolution, whereby spacetime is
foliated by null hypersurfaces that can be compactified towards \Scri
(see \cite{WinicourLRR} for a review).
This works well in the far field but characteristic foliations are ill
behaved in strong-field regions due to the formation of caustics in
the null congruences generating the hypersurfaces.
A compromise is Cauchy-characteristic matching, where a Cauchy
evolution in the interior is matched to a characteristic evolution in
an exterior region.
In \cite{Reisswig2009} this has recently been applied to
the {\it a posteriori} extraction of gravitational radiation from a
Cauchy evolution (with finite outer boundary) of a binary black hole 
coalescence.

A more efficient solution is a foliation of spacetime into
hyperboloidal hypersurfaces, which are everywhere spacelike but
approach future null infinity rather than spacelike infinity.
Such surfaces are not Cauchy surfaces---the foliation covers only the
region to the future of some initial hyperboloidal hypersurface but
not, for instance, the region around spacelike infinity.
This is not a serious disadvantage for the applications we have in
mind as we are mainly interested in computing the gravitational
radiation at \Scri and studying the late-time behaviour of perturbations.
Hyperboloidal foliations have the advantage of being as flexible as 
standard Cauchy foliations in the interior, so that a variety of gauge
conditions can be used that have been known to be successful in 
strong-field situations.
In \cite{Friedrich1983a} Friedrich reformulated the Einstein equations for
a conformally rescaled metric in such a way that they are manifestly
regular at \Scri.
This symmetric hyperbolic system is larger than the actual Einstein 
equations, e.g.~it contains evolution equations for the Weyl curvature.
For reviews of the theoretical development as well as 
numerical implementations, we refer the reader 
to \cite{FrauendienerLRR,Husa2002,Husa2003}.
Perhaps most relevant to the present work is an axisymmetric
reduction of the regular conformal field equations by Frauendiener
and Hein \cite{Frauendiener2002} that was applied to the numerical 
evolution of Minkowski spacetime and the boost-axisymmetric 
solutions of Bi\v{c}\'{a}k and Schmidt \cite{Bicak1989}.

Despite these successes, most recent work in numerical relativity has
used formulations of the Einstein equations (usually on truncated Cauchy
hypersurfaces) rather than the regular conformal field equations.
For this reason, it is worth investigating whether some formulation 
of the Einstein equations themselves can be adapted to hyperboloidal 
foliations reaching out to \Scri.
As for the regular conformal field equations, one applies a conformal
transformation to the metric in order to map \Scri to a finite
coordinate location in a compactified coordinate system, where the
conformal factor vanishes.
When written in terms of the conformal metric, the Einstein
equations contain inverse powers of the conformal factor,
terms that are singular at \Scri.
Numerical implementations must face the question of how to deal with
these formally singular terms, which may be potential sources of instabilities.

One proposal for solving the Einstein equations on hyperboloidal
slices was developed by Zengino\u{g}lu \cite{Zenginoglu2008}.
It combines a conformal transformation of the spacetime metric (where
the conformal factor can essentially be freely specified) with a
certain choice of gauge source functions for the Einstein equations in
(generalized) harmonic gauge \cite{Friedrich1985}.
Preliminary numerical results on spherically symmetric evolutions of
Schwarzschild spacetime were reported in section 2.5 
of \cite{ZenginogluPhD}.
However, it is not yet clear how the formally singular terms at
\Scri are to be evaluated numerically in more general situations.

Here we follow a different approach due to Moncrief that is based 
on an ADM-like \cite{Arnowitt1962} formulation of the Einstein 
equations with a constant-mean-curvature (CMC) slicing condition.
In \cite{Moncrief2009} we showed explicitly how the formally singular 
terms in the evolution equations at \Scri can be evaluated there in
terms of conformally regular geometric data.
The constraint equations that occur in this approach were solved
numerically in \cite{Buchman2009} and initial data on CMC slices 
describing various configurations of single and binary black holes were
constructed.
The present paper is concerned with the evolution problem.
We assume spacetime to be axisymmetric in order to speed up the code
so that different ways of treating \Scri numerically can be
experimented with.
We apply the ideas of \cite{Moncrief2009} to an earlier axisymmetric 
evolution scheme on Cauchy slices \cite{Rinne2008a}
(see also \cite{Garfinkle2001,Choptuik2003} for related schemes).
The objective of the paper is to show that long-term stable evolutions
are possible, including gravitational radiation.

The paper is organized as follows.
In section \ref{s:Formulation} we explain our notation and gauge
choices and obtain the constraint and evolution equations.
Particular emphasis is placed on how the formally singular evolution
equations can be regularized at \Scri.
In section \ref{s:NumericalMethod} we describe our numerical method
for solving these equations.
The numerical results are presented in section \ref{s:NumericalResults}.
As a first test problem we evolve Schwarzschild spacetime and
demonstrate the convergence of the numerical solution.
Next, a gravitational perturbation is included.
A regular expression at \Scri for the Bondi news function \cite{Bondi1962}
describing the outgoing gravitational radiation is derived and 
evaluated numerically.
We conclude and discuss possible directions for future work in 
section \ref{s:Concl}.


\section{Formulation}
\label{s:Formulation}


\subsection{Definitions and gauge choices}
\label{s:Defs}

We assume that spacetime is axisymmetric with Killing vector $\xi$.
For simplicity we also assume here that $\xi$ is hypersurface orthogonal.
Spherical polar coordinates $t,r,\theta,\phi$ are chosen such that
$\xi = \partial/\partial\phi$. 

The spacetime metric $\four{g}_{\alpha\beta}$ is written as
\begin{equation}
  \four{g}_{\alpha\beta} = \psi^{-2} \four{\gamma}_{\alpha\beta},
\end{equation}
where $\psi$ is the conformal factor and $\four{\gamma}_{\alpha\beta}$ 
the conformal metric.
Greek indices $\alpha,\beta,\ldots$ are spacetime indices ($t,r,\theta,\phi$).
The conformal factor $\psi$ is positive and approaches zero at future null
infinity, which we put at a fixed coordinate location $r=1$.
The conformal metric $\four{\gamma}_{\alpha\beta}$ is assumed to be
regular there.

We decompose the metric in ADM form as
\begin{equation}
  \four{g}_{\alpha\beta} \rmd x^\alpha \rmd x^\beta 
  = -\alpha^2 \rmd t^2 + g_{ij} (\rmd x^i +
  \beta^i \rmd t) (\rmd x^j + \beta^j \rmd t),
\end{equation}
where $g_{ij}$ is the spatial metric, $\alpha$ the lapse function
and $\beta^i$ the shift vector.
Latin indices $i,j,\ldots$ are spatial ($r,\theta,\phi$).
Similarly, we write the conformal spacetime metric as
\begin{equation}
  \label{e:Conformal4metric}
  \four{\gamma}_{\alpha\beta} \rmd x^\alpha \rmd x^\beta 
  = -\tilde \alpha^2 \rmd t^2 + \gamma_{ij} (\rmd x^i +
  \beta^i \rmd t) (\rmd x^j + \beta^j \rmd t),
\end{equation}
where $\gamma_{ij} = \psi^2 g_{ij}$ is the conformal spatial metric
and $\talpha = \psi \alpha$ the conformal lapse.

The spatial coordinates are chosen such that $\beta^\phi$ = 0 and the
conformal spatial metric takes the form
\begin{equation}
  \label{e:Conformal3metric}
  \gamma_{ij} \rmd x^i \rmd x^j = 
    \rme^{2\eta \sin\theta} (\rmd r^2 + r^2 \rmd \theta^2) 
    + r^2 \sin^2\theta \, \rmd \phi^2.
\end{equation}
This is sometimes called \emph{quasi-isotropic gauge}.
We note that this choice of gauge differs from the spatial harmonic 
gauge used in \cite{Moncrief2009}.\footnote{It is, in a sense, 
its two-dimensional analogue: if we define a two-dimensional metric 
$H_{AB}$ by the part of the line element \eref{e:Conformal3metric} 
orthogonal to the Killing vector,
$H_{AB} \rmd x^A \rmd x^B = \rme^{2\eta \sin\theta} 
(\rmd r^2 + r^2 \rmd \theta^2)$,
then $V^C \equiv H^{AB} (\Gamma^C{}_{AB} - \Gammaz^C{}_{AB}) = 0$,
where $\Gamma^C{}_{AB}$ are the Christoffel symbols of $H_{AB}$ and 
$\Gammaz^C{}_{AB}$ are the Christoffel symbols of the flat metric 
($H_{AB}$ with $\eta=0$). [The author thanks V.~Moncrief for pointing
this out.]}

The extrinsic curvature is defined by
\begin{equation}
  \label{e:ExtrCurv}
  \partial_t g_{ij} = -2\alpha K_{ij} + \Lie_\beta g_{ij},
\end{equation}
where $\Lie$ denotes the Lie derivative.
Preservation of the conditions $g_{\theta\theta} = r^2 g_{rr}$ and
$g_{r\theta} = 0$ under the evolution equation \eref{e:ExtrCurv} implies
\begin{eqnarray}
  \label{e:Shift1r}
  \beta^r{}_{,r} - \beta^\theta{}_{,\theta} - r^{-1} \beta^r 
  = \alpha (K^r{}_r - K^\theta{}_\theta) \equiv \talpha \tU,\\
  \label{e:Shift1theta}
  \beta^r{}_{,\theta} + r^2 \beta^\theta{}_{,r} = 2 \alpha K^r{}_\theta
  \equiv 2 \talpha \tKrth.
\end{eqnarray}
Similarly as in \cite{Rinne2008a} we have defined the following 
components of the extrinsic curvature,
\begin{equation}
  \label{e:DefUKrth}
  K^r{}_r - K^\theta{}_\theta \equiv \psi \tU, \qquad
  K^r{}_\theta = r^{-2} K^\theta{}_r \equiv \psi \tKrth ,
\end{equation}
and, motivated by regularity concerns on the axis of 
symmetry \cite{Rinne2008a,Garfinkle2001,Choptuik2003},
we also set
\begin{equation}
  \label{e:DefW}
  K^\theta{}_\theta - K^\phi{}_\phi \equiv \psi \sin\theta \, \tW.
\end{equation}
It is easy to see that the quantities $\tU$, $\tKrth$
and $\tW$ defined in \eref{e:DefUKrth} and \eref{e:DefW} 
have the same conformal weight as the traceless momentum 
\begin{equation}
  \pi^{\mathrm{tr}\, ij} = \sqrt{g} (K^{ij} - \third g^{ij} K_l{}^l)
\end{equation}
used in \cite{Moncrief2009}, where $g$ denotes the determinant of $g_{ij}$.
Note also that because $\tKrth$, $\tW$ and $\tU$
are either off-diagonal or differences of diagonal components 
of the extrinsic curvature, they can in fact be regarded as the three 
independent components of the \emph{traceless} part of the extrinsic
curvature in our case.

As in \cite{Moncrief2009} we require the time slices to have constant
mean curvature,
\begin{equation}
  \label{e:CMCdef}
  g^{ij} K_{ij} \equiv -K = \const .
\end{equation}
Note the slightly awkward sign in the definition of the
constant $K$; with our sign convention for the extrinsic curvature
\eref{e:ExtrCurv} we require $K$ to be \emph{positive} so that the slices
reach \emph{future} null infinity.
Preservation of \eref{e:CMCdef} under the Einstein evolution equation for
the extrinsic curvature yields the following elliptic equation for the
conformal lapse,
\begin{eqnarray}
  \label{e:CMClapse}
  \fl 0 = \talpha_{,rr} + 2r^{-1} \talpha_{,r} + r^{-2}
  (\talpha_{,\theta\theta} + \cs \talpha_{,\theta})\nonumber\\
  - \half s \talpha (\eta_{,rr} + r^{-1} \eta_{,r} + r^{-2}
  \eta_{,\theta\theta} + 2 r^{-2} \cs \eta_{,\theta} - r^{-2} \eta)\nonumber\\
  + \tfrac{3}{2} \talpha \left[ P_r(P_r - 2 \tA_r) + r^{-2}
    P_\theta(P_\theta - 2 \tA_\theta) \right] 
  - \tfrac{1}{6} \psi^{-2} \talpha \epluss K^2\nonumber\\
  - \tfrac{5}{2}\talpha \epluss \left[ \third(\tilde U + \half s\tW)^2
  + \fourth (s\tilde W)^2 + r^{-2} (\tKrth)^2 \right] . 
\end{eqnarray}
Here and in the following we use the abbreviations
\begin{equation}
  s \equiv \sin\theta, \quad c \equiv\cos\theta,\quad
  P_B \equiv \psi^{-1} \psi_{,B}, \quad \tA_B \equiv \talpha^{-1} \talpha_{,B},
\end{equation}
upper-case Latin indices $A,B,\ldots$ ranging over $r$ and $\theta$.


\subsection{Constraint and evolution equations}

As usual in ADM-like reductions, the Einstein equations split into
constraint and evolution equations.
The Hamiltonian constraint is
\begin{eqnarray}
  \label{e:Hamcons}
  \fl 0 = \psi_{,rr} + 2 r^{-1} \psi_{,r} + r^{-2} (\psi_{,\theta\theta} 
  + \cs \psi_{,\theta})
  - \tfrac{3}{2} \psi^{-1} \left[(\psi_{,r})^2 + r^{-2}
    (\psi_{,\theta})^2 \right]\nonumber\\
  - \half s \psi (\eta_{,rr} + r^{-1} \eta_{,r} + r^{-2}
  \eta_{,\theta\theta} + 2 r^{-2} \cs \eta_{,\theta} - r^{-2} \eta)\nonumber\\
  - \half \psi \epluss \left[ \third(\tilde U + \half s\tW)^2
  + \fourth (s\tilde W)^2 + r^{-2} (\tKrth)^2 \right]\nonumber\\
  + \tfrac{1}{6} \psi^{-1} \epluss K^2 .
\end{eqnarray}
The momentum constraints are
\begin{eqnarray}
  \label{e:Momconsr}
  \fl 0 = \cC^r \equiv 
  \tfrac{2}{3} \tU_{,r} + \third s \tW_{,r} + r^{-2} \tKrth_{,\theta}
  + r^{-2} \tKrth (\cs + 2 c\eta + 2 s\eta_{,\theta} - 2P_\theta)\nonumber\\
  + \tU(s\eta_{,r} - \tfrac{4}{3} P_r + 2 r^{-1})
  + s \tW(r^{-1} - \tfrac{2}{3} P_r),\\
  \label{e:Momconstheta}
  \fl 0 = \cC^\theta \equiv
  -\third \tU_{,\theta} + \third s \tW_{,\theta} + \tKrth_{,r}
  + 2 \tKrth(s\eta_{,r} - P_r + r^{-1})\nonumber\\
  + \tU(-c\eta - s\eta_{,\theta} + \tfrac{2}{3}P_\theta)
  + \tW(\tfrac{4}{3}c - \tfrac{2}{3} s P_\theta).
\end{eqnarray}

We have the following evolution equations,
\begin{eqnarray}
  \label{e:dtpsi}
  \fl \psi_{,t} = \beta^r \psi_{,r} + \beta^\theta \psi_{,\theta}
  - (\cs \beta^\theta + r^{-1} \beta^r) 
  - \third \talpha (K + \psi (\tU + 2 s \tW),\\
  \label{e:dteta}
  \fl \eta_{,t} = \beta^r \eta_{,r} + \beta^\theta
  \eta_{,\theta} + \cs \beta^\theta \eta + (s^{-1}
    \beta^\theta)_{,\theta} - \talpha \tW,\\
  \label{e:dtW}  
  \fl \tW_{,t} = \beta^r \tW_{,r} + \beta^\theta \tW_{,\theta}
  + (2\cs \beta^\theta + r^{-1} \beta^r)\tW \nonumber\\
  + \eminuss r^{-2} \left[ -(s^{-1} \talpha_{,\theta})_{,\theta}
    + 2 \talpha\psi^{-1} (s^{-1} \psi_{,\theta})_{,\theta} \right]
  \nonumber\\
  - \talpha\eminuss \big\{ \eta_{,rr} + 2r^{-1} \eta_{,r} 
    + r^{-2} [\eta_{,\theta\theta} - \eta + c (s^{-1}\eta)_{,\theta}]
  \nonumber\\ \qquad 
  + (\tA_r - 2 P_r) \eta_{,r}
  - r^{-2} s^{-1} (\tA_\theta - 2 P_\theta)(s \eta_{,\theta} + c \eta) 
  \big\} \nonumber\\
  -2s^{-1}  \tKrth \beta^\theta{}_{,r}
  + \tfrac{2}{3} \talpha s^{-1} \left[ s\tW(s\tW + \half \tU -
      \psi^{-1} K) + 3 r^{-2} (\tKrth)^2 \right]\\
  \label{e:dtKrth}
  \fl \tKrth_{,t} = \beta^r \tKrth_{,r} + \beta^\theta
  \tKrth_{,\theta} + \cs \beta^\theta \tKrth
  + \eminuss (-\talpha_{,r\theta} + 2 \talpha \psi^{-1} \psi_{,r\theta})\nonumber\\
  + \talpha \eminuss \big[ (\tA_r - 2 P_r + r^{-1}) (s \eta_{,\theta} + c\eta)
  + (\tA_\theta - 2 P_\theta) (s\eta_{,r} + r^{-1}) \nonumber \\ \qquad
  + c \eta_{,r}  \big]
  + \tfrac{2}{3} \talpha\tKrth(s\tW -\psi^{-1} K + 2 \tU) - r^2
  \beta^\theta{}_{,r} \tU,\\
  \label{e:dtU}
  \fl \tU_{,t} = \beta^r \tU_{,r} + \beta^\theta \tU_{,\theta} +
  \tU(\cs\beta^\theta + r^{-1} \beta^r)\nonumber\\
  + \eminuss[-\talpha_{,rr} + r^{-2} \talpha_{,\theta\theta} +
  2\talpha\psi^{-1}(\psi_{,rr} -r^{-2}\psi_{,\theta\theta})]\nonumber\\
  + \talpha\eminuss [(\tA_r - 2 P_r)(2s\eta_{,r} + r^{-1})
  + 2 s r^{-1} \eta_{,r} \nonumber\\ \qquad 
  - 2 r^{-2} (\tA_\theta - 2 P_\theta)(s\eta_{,\theta} + c\eta + \tfrac{c}{s}) ]
  \nonumber\\
  + \third\talpha\tU(2s\tW - 2 \psi^{-1} K + \tU) 
  + 4 \tKrth (\beta^\theta{}_{,r} - \talpha r^{-2} \tKrth).
\end{eqnarray}

We observe that all the equations are manifestly regular at the axis
of symmetry $\theta=0$ when the parities of the fields with
respect to $\theta=0$ are taken into account (table \ref{t:Parities}).
These can be inferred from the general behaviour of regular
axisymmetric tensor fields \cite{Rinne2005}.

\begin{table}
  \caption{\label{t:Parities}
    Angular parities of the fundamental variables: even ($+$) or odd ($-$).
    The parities about $\theta=0$ follow from regularity on axis,
    those about $\theta=\pi/2$ from the reflection symmetry we impose.
}
  \centering
  \begin{tabular}{l|c|c|c|c|c|c|c|c}
    Variable & $\talpha$ & $\beta^r$ & $\beta^\theta$ & $\psi$ 
      & $\eta$ & $\tW$ & $\tKrth$ & $\tU$ \\\hline
    $\theta=0$ & $+$ & $+$ & $-$ & $+$ & $-$ & $-$ & $-$ & $+$\\
    $\theta=\pi/2$ & $+$ & $+$ & $-$ & $+$ & $+$ & $+$ & $-$ & $+$ 
  \end{tabular}  
\end{table}


\subsection{Partially constrained evolution and hyperbolicity}
\label{s:Hyperbolicity}

We determine the conformal factor $\psi$ by solving the momentum
constraint \eref{e:Hamcons} rather than the evolution equation \eref{e:dtpsi}.
On the other hand, the extrinsic curvature variables $\tW, \tKrth$ and $\tU$
are evolved using the evolution equations \eref{e:dtW}--\eref{e:dtU} and
the momentum constraints \eref{e:Momconsr} and \eref{e:Momconstheta} are
not solved explicitly.
Hence we adopt a partially constrained evolution scheme.

Of course the constraints are preserved by the evolution
equations, which imply the following subsidiary system,
\begin{eqnarray}
  \cC^t{}_{,t} &\simeq& \beta^r \cC^t{}_{,r} + \beta^\theta \cC^t{}_{,\theta}
  - \half \talpha\psi(\cC^r{}_{,r} + r^{-2} \cC^\theta{}_{,\theta}),\\
  \cC^r{}_{,t} &\simeq& \beta^r \cC^r{}_{,r} + \beta^\theta \cC^r{}_{,\theta}
  + \tfrac{4}{3} \eminuss \talpha \psi^{-1} \cC^t{}_{,r},\\
  \cC^\theta{}_{,r} &\simeq& \beta^r \cC^\theta{}_{,r} 
  + \beta^\theta \cC^\theta{}_{,\theta}
  + \tfrac{4}{3} \eminuss \talpha \psi^{-1} \cC^t{}_{,\theta},\\
\end{eqnarray}
where the symbol $\simeq$ signifies that only the principal parts have
been displayed, and the remaining terms are homogeneous in the constraints.
If the Hamiltonian constraint is enforced exactly, $\cC^t = 0$, then
the remaining evolution equations for $\cC^r$ and $\cC^\theta$ are
clearly hyperbolic (they are advection equations along the shift).
If the Hamiltonian constraint is not enforced then the full
constraint evolution system is \emph{not} hyperbolic (the
characteristic speeds 
\begin{equation}
  \lambda^r = \beta^r \pm \sqrt{\tfrac{2}{3}} \, \eminus 
  \talpha \, \rmi , \qquad
  \lambda^\theta = \beta^\theta \pm \sqrt{\tfrac{2}{3}} \, r^{-1} \eminus 
  \talpha \, \rmi 
\end{equation}
are complex).
This might be cured by adding multiples of the Hamiltonian constraint
to the evolution equations; however, we have not found such a constraint
addition that both yields a hyperbolic constraint evolution system and
is compatible with regularity of the main evolution equations on the
axis of symmetry.
Hence a free evolution scheme that solves none of the constraint
equations during the evolution does not appear to be feasible.
However, our formulation includes three elliptic
equations for the gauge variables $\talpha$ and $\beta^A$ 
anyway and solving one additional elliptic equation for the
conformal factor is not overly expensive.

If we assume that $\talpha, \beta^A$ and $\psi$ are given by the
solution of the elliptic equations, then the pair of evolution equations
\eref{e:dteta} and \eref{e:dtW} forms a wave equation to principal
parts,
\begin{equation}
  (\partial_t - \beta^A \partial_A)^2 \eta \simeq
  \talpha^2 \eminuss (\eta_{,rr} + r^{-2} \eta_{,\theta\theta}),
\end{equation}
and the evolution equations \eref{e:dtKrth} and \eref{e:dtU} are
advection equations along the shift to principal parts.
Hence the evolution equations are hyperbolic.
A rigorous proof of the well posedness of the present mixed
elliptic-hyperbolic hyperboloidal initial value problem remains to be
carried out.
For a similar mixed elliptic-hyperbolic formulation on a Cauchy
foliation, such a proof was given in in \cite{Andersson2003}.


\subsection{Regularity at future null infinity}
\label{s:Regularity}

The CMC slicing condition \eref{e:CMClapse}, the constraint equations
\eref{e:Hamcons}--\eref{e:Momconstheta} and the evolution equations
for the extrinsic curvature \eref{e:dtW}--\eref{e:dtU} are formally
singular at \Scri, where the conformal factor $\psi$ vanishes.
Our elliptic solver (section \ref{s:MG}) is well adapted to
such degenerate elliptic equations.
However, the right-hand sides of the evolution equations cannot be
evaluated at \Scri in their present form.
In \cite{Moncrief2009} two methods for obtaining regular forms of the
evolution equations at \Scri were presented: the first based on
Taylor expansions, the second on an argument due to Penrose \cite{Penrose1965}.
For the numerical results presented in section \ref{s:NumericalResults} 
we followed the second approach but we have also implemented the first
approach, as discussed further below.

Penrose showed that provided certain smoothness assumptions hold 
(in particular, the conformal metric must be $C^3$ up to the
boundary), the conformal Weyl tensor (i.e., corresponding to the 
conformal metric $\four{\gamma}_{\alpha\beta}$) vanishes at \Scri.
This implies \cite{Moncrief2009} that the electric part of the
\emph{physical} Weyl tensor (i.e., corresponding to $\four{g}_{\alpha\beta}$)
also vanishes at \Scri,
\begin{equation}
  E_{\alpha\gamma} = n^\beta n^\delta
  C_{\alpha\beta\gamma\delta}[\four{g}] \hateq 0 ,
\end{equation}
where $n^\alpha$ is the unit normal to the $t=\const$ slices and
$\hateq$ denotes equality at \Scri.
It is straightforward to compute the Weyl tensor for the metric
$\four{\gamma}_{\alpha\beta}$ defined in
\eref{e:Conformal4metric} and \eref{e:Conformal3metric}.
Any time derivatives are substituted using the evolution equations.
(Here a computer algebra program is very useful; we used 
REDUCE \cite{ReduceManual}.)
Setting $E_{\alpha\gamma} \hateq 0$ immediately gives regular
expressions for the formally singular terms in
\eref{e:dtW}--\eref{e:dtU}.
Thus we obtain the following manifestly regular evolution equations at \Scri,
\begin{eqnarray}
  \label{e:dtWreg}  
  \fl \tW_{,t} \hateq \beta^r \tW_{,r} + \beta^\theta \tW_{,\theta}
  + (2\cs \beta^\theta + r^{-1} \beta^r)\tW 
  - \eminuss r^{-2} (s^{-1} \talpha_{,\theta})_{,\theta}
  \nonumber\\
  + \talpha\eminuss \big\{ \eta_{,rr} + 2r^{-1} \eta_{,r} 
    + r^{-2} [\eta_{,\theta\theta} - \eta + c (s^{-1}\eta)_{,\theta}]
  \nonumber\\ \qquad 
  - A_r \eta_{,r} + r^{-2} s^{-1} A_\theta (s\eta_{,\theta} + c\eta)
  \big\} \nonumber\\
  -2s^{-1}  \tKrth \beta^\theta{}_{,r}
  + \talpha [ 4s^{-1}r^{-2}(\tKrth)^2 - \tU\tW],\\
  \fl \tKrth_{,t} \hateq \beta^r \tKrth_{,r} + \beta^\theta
  \tKrth_{,\theta} + \cs \beta^\theta \tKrth
  - \eminuss \talpha_{,r\theta} \nonumber\\
  + \talpha \eminuss \big[ (\tA_r - r^{-1}) (s \eta_{,\theta} + c\eta)
  + \tA_\theta (s\eta_{,r} + r^{-1}) - c \eta_{,r}  \big]\nonumber\\
  + 2 \talpha\tKrth(s\tW + \tU) - r^2 \beta^\theta{}_{,r} \tU,\\
  \label{e:dtUreg}
  \fl \tU_{,t} \hateq \beta^r \tU_{,r} + \beta^\theta \tU_{,\theta} +
  \tU(\cs\beta^\theta + r^{-1} \beta^r)
  + \eminuss(-\talpha_{,rr} + r^{-2} \talpha_{,\theta\theta})\nonumber\\
  + \talpha\eminuss [\tA_r (2s\eta_{,r} + r^{-1})
  - 2 r^{-2} (\tA_\theta - \cs) (s\eta_{,\theta} + c\eta)
  - 2 s r^{-1} \eta_{,r}] \nonumber\\
  + \talpha\tU(2s\tW + \tU) 
  + 4 \tKrth (\beta^\theta{}_{,r} - \talpha r^{-2} \tKrth).
\end{eqnarray}

The other approach to regularity at \Scri discussed in \cite{Moncrief2009}
is based on an expansion of all the fields in (finite) Taylor series with
respect to radius $r$ about \Scri.
Substituting these expansions in the singular constraint equations,
one obtains conditions on the fields and their radial derivatives that
can be used in order to evaluate the formally singular terms in the
evolution equations.
This approach is more general in that it does not require Penrose's
smoothness assumption.

As necessary conditions for a regular evolution at \Scri, we obtained
in \cite{Moncrief2009} a set of conditions that had been discovered
in \cite{Andersson1992}, namely that the shear of \Scri
and the radial components $\pi^{\mathrm{tr} \, ri}$ of the traceless
momentum vanish.
In the axisymmetric formulation we consider here, these conditions read
\begin{equation}
  \label{e:RegularityConditions}
  \tW \hateq -\rme^{-s\eta} \eta_{,r}, \qquad
  \tU \hateq \half \rme^{-s\eta} s \eta_{,r}, \qquad
  \tKrth \hateq 0.
\end{equation}
One might be tempted to impose these conditions as Dirichlet
conditions on the extrinsic curvature at \Scri, thus avoiding to
evaluate the singular evolution equations there.
However, numerical evolutions with these Dirichlet boundary conditions were
found to be unstable.
This is perhaps not surprising because at \Scri all the fields are
purely outgoing and hence imposing boundary conditions there is ill posed
(although of course the conditions \eref{e:RegularityConditions} 
are satisfied for a \emph{given} solution of the field equations).
It is worthwhile to observe though, as we already did in \cite{Moncrief2009}, 
that the evolution equations \eref{e:dtW}--\eref{e:dtU} contain 
remarkable damping terms that cause potential violations of the 
regularity conditions \eref{e:RegularityConditions} to decay 
exponentially in the neighbourhood of \Scri:
\begin{equation}
  u_{,t} = \ldots - \tfrac{2}{3} \talpha \psi^{-1} K u
\end{equation}
for each $u \in \{ \tW, \tKrth, \tU \}$.
The term has the ``right sign'' because $K>0$.

Instead of imposing the regularity conditions \eref{e:RegularityConditions} 
directly, however, we can extract more information from the Taylor
expansions that allows us to evaluate the singular terms in the
evolution equations directly \cite{Moncrief2009}.
The resulting modified evolution equations at \Scri are detailed for
the current axisymmetric scheme in \ref{s:TaylorReg}.
Their main difference to \eref{e:dtWreg}--\eref{e:dtUreg} lies in the
presence of terms involving first radial derivatives of the extrinsic
curvature at \Scri (cf.~equation \eref{e:dtWreg2}).
We found this version of the evolution equations at \Scri to be
numerically stable as well, and the results agree with those in section
\ref{s:NumericalResults} within numerical error (see \ref{s:TaylorReg}
for details).

It is worth noting here that the quasi-isotropic spatial gauge we use 
differs from the spatially harmonic gauge of \cite{Moncrief2009}.
Nevertheless, we were able to apply the same procedure for evaluating
the formally singular terms at \Scri.
This provides some substance to the claim in \cite{Moncrief2009} that
the precise form of the spatial gauge conditions is inessential for
the method to be applicable.


\section{Numerical method}
\label{s:NumericalMethod}

Having obtained a set of gauge conditions, constraints and evolution 
equations, we now describe how we solve them numerically.
First we summarize the evolution scheme, i.e.~which variables are
evolved using which equations.
Next we specify our choice for the computational domain and the
discretization of the partial differential equations.
We then focus on the two main aspects of the code, the time
integration scheme and the elliptic solver.
The code has been written in C++.
Unless otherwise indicated all routines have been developed from scratch.


\subsection{Evolution scheme}

Our fundamental variables are $\talpha,\beta^A,\psi,\eta,\tW,\tKrth$
and $\tU$.
The variables $\eta,\tW,\tKrth$ and $\tU$ are evolved using the evolution
equations \eref{e:dteta}--\eref{e:dtU}.
At each time step we solve the CMC slicing condition \eref{e:CMClapse}
for $\talpha$, the quasi-isotropic gauge conditions
\eref{e:Shift1r} and \eref{e:Shift1theta} for $\beta^A$, 
and the Hamiltonian constraint \eref{e:Hamcons} for $\psi$.
Details on how the quasi-isotropic gauge conditions are treated are given in
\ref{s:Shift}.
The momentum constraints are only solved initially (\ref{s:Momcons}).


\subsection{Domain and discretization}

For the applications considered in this paper, the spatial numerical
domain is taken to be the quarter of an annulus, 
$\rmin \leqslant r \leqslant 1$ and $0 \leqslant \theta \leqslant \pi/2$.
The inner boundary at $r=\rmin$ will be taken to lie within a black
hole event horizon (black hole excision), 
and the outer boundary at $r=1$ corresponds to \Scri.
The axis of symmetry is at $\theta = 0$, and we impose an additional 
reflection symmetry about the plane $\theta=\pi/2$.

The fields typically have the steepest gradients close to the black
hole and hence it is advisable to use a non-uniform radial grid in
order to provide more numerical resolution towards the inner boundary.
We introduce a quadratic map
\begin{equation}
  \label{e:RadialMap}
  r(x) = Q x^2 + (1-\rmin-Q)x + \rmin ,
\end{equation}
where $0\leqslant Q <1$ is a constant.
This map satisfies $r(0) = \rmin$ and $r(1) = 1$.
We use a uniform grid in $x$, i.e.~we place grid points at 
$x_i = i/N_r$, $0\leqslant i \leqslant N_r$,
where $N_r$ is the number of radial grid points.
Note in particular that there are grid points right at the inner
boundary and at \Scri.
In the angular direction we use a uniform staggered grid:
$\theta_j = \tfrac{\pi}{2} (j-\half)/N_\theta$, 
$1\leqslant j \leqslant N_\theta$.
Two layers of ghost points are added on either side, corresponding to
$j=-1,0$ and $j=N_{\theta+1},N_{\theta+2}$.
Values at these ghost points are set according to the angular 
parities of the various fields, which are listed in table \ref{t:Parities}.
E.g.~for a field $u$ that is odd about $\theta=0$ we set 
$u_{i0} = -u_{i,1}$ and $u_{i,-1} = -u_{i2}$, and for an even field we set
$u_{i0} = u_{i,1}$ and $u_{i,-1} = u_{i2}$.
Here $u_{ij}$ denotes the value of $u$ at the grid point with indices
$(i,j)$.

The spatial derivatives are discretized using fourth-order accurate finite
differences. 
One-sided differences are used in the radial direction at the boundaries;
everywhere else centred differences are used (see \ref{s:FD} for details).


\subsection{Time integration}

A fourth-order Runge-Kutta method is used in order to integrate the
evolution equations forward in time.
At each substep of this method, the elliptic equations are solved as
described below in section \ref{s:MG}.
Since all the characteristics point towards the exterior of the domain
both at the inner excision boundary and at \Scri, no boundary conditions are
imposed there---the evolution equations are discretized using one-sided
finite differences.
At \Scri the evolution equations \eref{e:dtW}--\eref{e:dtU} for the
extrinsic curvature are replaced with their regularized versions
\eref{e:dtWreg}--\eref{e:dtUreg}.
Kreiss-Oliger dissipation \cite{Kreiss1973}
is added in order to ensure stability (\ref{s:FD}).
No modifications to the dissipation operators or special smoothing
operations at \Scri have been required.


\subsection{Elliptic solver and boundary conditions}
\label{s:MG}

The elliptic equations are solved by means of a multigrid 
method \cite{Brandt1977,BriggsMG} using Full Approximation Storage (FAS).
Because of the singular nature of the equations at \Scri ($r=1$), we
have found it necessary to use line relaxation in the radial
direction, i.e.~all points on a line $\theta=\const$ are solved
simultaneously.
With our fourth-order finite-difference discretization
this requires solving a penta-diagonal linear system of equations, for which
a fast algorithm exist (we use the routines \texttt{bandec} and
\texttt{banbks} in \cite{NumericalRecipes}).
For the Hamiltonian constraint \eref{e:Hamcons} an outer
Newton-Raphson iteration is employed for each such line solve; 
all remaining elliptic equations are linear.
The lines $\theta=\const$ are then traversed and updated in ascending order
$j=1,2,\ldots,N_\theta$ (Gauss-Seidel relaxation in the $\theta$-direction).
We have found this method to be very robust and effective, leading to a
reduction of the residual of about an order of magnitude per W-cycle.
No problems with convergence at the singular boundary were encountered.

The elliptic equations require boundary conditions at the inner
$r=\rmin$ and outer $r=1$ boundary (in addition to the parity 
boundary conditions at $\theta=0,\pi/2$), which we shall now specify.

At the outer boundary (\Scri) we take the angular shift to vanish,
$\beta^\theta \hateq 0$.
The conformal factor vanishes there by definition, $\psi \hateq 0$.
Preservation of this condition under the evolution equation
\eref{e:dtpsi} yields a boundary condition for the conformal lapse,
\begin{equation}
  \label{e:LapseBC}
  \talpha \hateq -\eplus \beta^r. 
\end{equation}
We remark that these boundary conditions for the lapse and shift 
ensure that the normal $n^\alpha$ to the time slices is null at \Scri.
The radial shift $\beta^r$ appearing in \eref{e:LapseBC} cannot be chosen 
freely but is determined by the quasi-isotropic gauge conditions.
After each Gauss-Seidel relaxation sweep we integrate \eref{e:Shift1theta}
along the boundary to determine $\beta^r$ there.
The integration constant can be fixed e.g.~by requiring that the 
average of $\beta^r$ along the boundary agree with the value obtained from
the exact Schwarzschild solution that we consider in 
section \ref{s:NumericalResults}.

At the inner excision boundary, we also take $\beta^\theta$ to vanish, 
and we freeze $\talpha$ to the value corresponding to the exact
Schwarzschild solution.
However, the conformal factor $\psi$ \emph{cannot} be chosen freely.
This is because $\psi$ obeys an evolution equation with
outgoing characteristic speed at the excision boundary, 
equation \eref{e:dtpsi}.
In order to determine the correct value of $\psi$ there,
we keep a copy of $\psi$ that is evolved using \eref{e:dtpsi}.
At the end of each timestep this copy of $\psi$ is reset to
agree with the solution of the Hamiltonian constraint.
This procedure was found to be crucial for ensuring that the \emph{momentum}
constraints are preserved during the numerical evolution.


\section{Numerical results}
\label{s:NumericalResults}

In this section, we apply our code to the evolution of a (perturbed)
Schwarzschild black hole.
First we focus on the unperturbed Schwarzschild solution and
demonstrate that long-term stable and convergent evolutions can be obtained.
Next we include a small gravitational perturbation. 
The outgoing gravitational radiation as represented by the Bondi news
function is computed at \Scri.

\subsection{Schwarzschild spacetime}
\label{s:Schw}

First we describe how the exact solution is computed, which is needed
both for the boundary conditions of some of the elliptic equations 
(section \ref{s:MG}) and in order to compare with the numerical solution
during the evolution.
The Schwarzschild solution in CMC slicing was derived in
\cite{Brill1980,Malec2003}.
Its line element is
\begin{equation}
   ds^2 = -\left(1-\frac{2M}{\bar r}\right) \rmd t^2 + \frac{1}{f^2} \,
   \rmd {\bar r}^2 - \frac{2a}{f} \, dt \, \rmd \bar r 
   + {\bar r}^2 (\rmd\theta^2 + \sin^2\theta \, \rmd\phi^2),
\end{equation}
where
\begin{equation}
   f(\bar r) = \left(1-\frac{2M}{\bar r} + a^2\right)^{1/2}, \qquad
   a(\bar r) = \frac{K \bar r}{3} - \frac{C}{{\bar r}^2}\, ,
\end{equation}  
and $M$ (mass), $K$ (mean curvature) and $C$ are constants.

As in \cite{Buchman2009} we introduce a new radial coordinate $r$ (the
coordinate used in our evolution scheme) by requiring the spatial
metric to be conformally flat in the new coordinates. 
Hence $r$ obeys 
\begin{equation}
  \frac{\rmd r}{\rmd \bar r} = \frac{r}{\bar r f(\bar r)}.
\end{equation}
This ordinary differential equation is integrated numerically to 
high precision using Mathematica.
The boundary condition is that $r\rightarrow 1$ as $\bar r \rightarrow
\infty$. 

The metric variables defined in section \ref{s:Defs} are then given by
\begin{equation}
  \psi = \frac{r}{\bar r}, \quad
  \talpha = \frac{rf}{\bar r}, \quad
  \beta^r = -\frac{ra}{\bar r}, \quad
  \beta^\theta = 0, \quad
  \eta = 0.
\end{equation}
The extrinsic curvature variables are found to be
\begin{equation}
  \tU = -\frac{3C}{{\bar r}^2 r}, \quad 
  \tW = 0, \quad
  \tKrth = 0,
\end{equation}
and the vector $V^A$ introduced in order to solve the momentum
constraints initially (\ref{s:Momcons}) is
\begin{equation}
  V^r = \frac{C}{r^2}, \quad 
  V^\theta = 0.
\end{equation}

For the numerical evolutions presented in this paper we choose $M=1$,
$K=1/2$ and $C=2$ (although we have obtained stable evolutions across
a wide range of parameter values).
The inner boundary is placed at $\rmin = 0.05$, just inside the
horizon at $\bar r=2 \Leftrightarrow r=0.0635$.

We run the simulation with two different numerical resolutions
$(N_r,N_\theta)=(64,8)$ and $(128,16)$. 
The radial grid is nonuniform with $Q=3/4$ in \eref{e:RadialMap} in
both cases.

The time step $\Delta t$ is limited by the Courant-Friedrichs-Lewy 
condition
\begin{equation}
  \Delta t < \min \left( \frac{\Delta r}{|v_\pm^r|}, \frac{\Delta
      \theta}{|v_\pm^\theta|} \right),
\end{equation}
where $\Delta r$ is the grid spacing and $v^r_\pm$ are the incoming and
outgoing characteristic speeds of the evolution system 
\eref{e:dtpsi}--\eref{e:dtU} in the $r$-direction 
(similarly in the $\theta$-direction),
and the minimum is taken over all grid points.
For our choice of parameters the time step is constrained by the
outgoing radial characteristic speed at \Scri, 
\begin{equation}
  v^r_+ = -\beta^r + \eminus\talpha \hateq \frac{2K}{3}
\end{equation}
for the Schwarzschild solution.
The time step is taken to be about $3/4$ of the maximum allowed value
computed in this way: for the lower resolution we use $\Delta t = 0.06$.

\begin{figure}
\smallskip
\centerline{\includegraphics[scale=0.24]{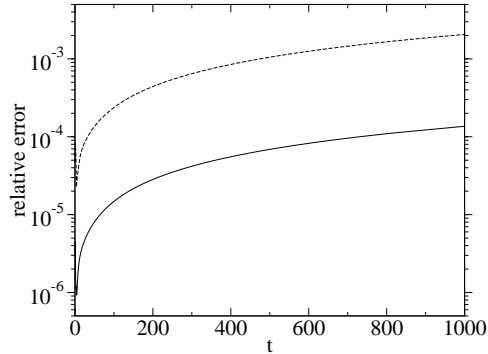}} 
\caption{\label{f:Schw} 
  Relative error of the numerical solution with respect to the exact
  solution for an evolution of Schwarzschild spacetime, computed as
  described in the text.
  Two different numerical resolutions are shown, 
  $(N_r,N_\theta) = (64,8)$ (dashed line) and $(128,16)$ (solid line).
}
\end{figure}

During the evolution, we compute the difference of the numerical
solution with respect to the exact solution for each of the evolved
variables and take its discrete $L_2$ norm.
This is divided by the norm of the exact solution for the
corresponding variable whenever the latter does not vanish.
Of these results a vector $2$-norm (square root of the sum of the squares) 
is taken.
The resulting quantity represents a measure of the overall relative
error of the numerical solution as a function of time and is plotted
in figure \ref{f:Schw}.
The error grows only linearly with time.
It decreases by a factor of about $15$ on average as the resolution is
doubled, close to the expected convergence factor of $2^4 = 16$ for a
fourth-order accurate scheme.
The main contribution to the error comes from the region close to the
black hole horizon where the fields have the steepest gradients.
The numerical solution remains smooth all the way up to \Scri during the 
entire evolution.

\subsection{Perturbed Schwarzschild spacetime}

Next, we include a small gravitational perturbation in our evolutions.
This is done by taking the variable $\eta$, which vanishes for the
exact Schwarzschild solution, to be a Gaussian initially,
\begin{equation}
  \eta = A \sin\theta \; \exp \left[-\frac{(r-r_c)^2}{2 \sigma^2}\right].
\end{equation}
The factor $\sin\theta$ has been included for parity reasons 
(table \ref{t:Parities}).
We choose $A=10^{-4}$, $r_c=0.5$ and $\sigma=0.05$.
We stress that these initial data are evolved with the full nonlinear
Einstein equations.
The linearized problem has been studied extensively; in particular, 
numerical evolutions of the Bardeen-Press equation \cite{Zenginoglu2009} 
and the Regge-Wheeler-Zerilli equations \cite{Zenginoglu2009b} 
have been performed on hyperboloidal slices.

\begin{figure}
\includegraphics[scale=0.24]{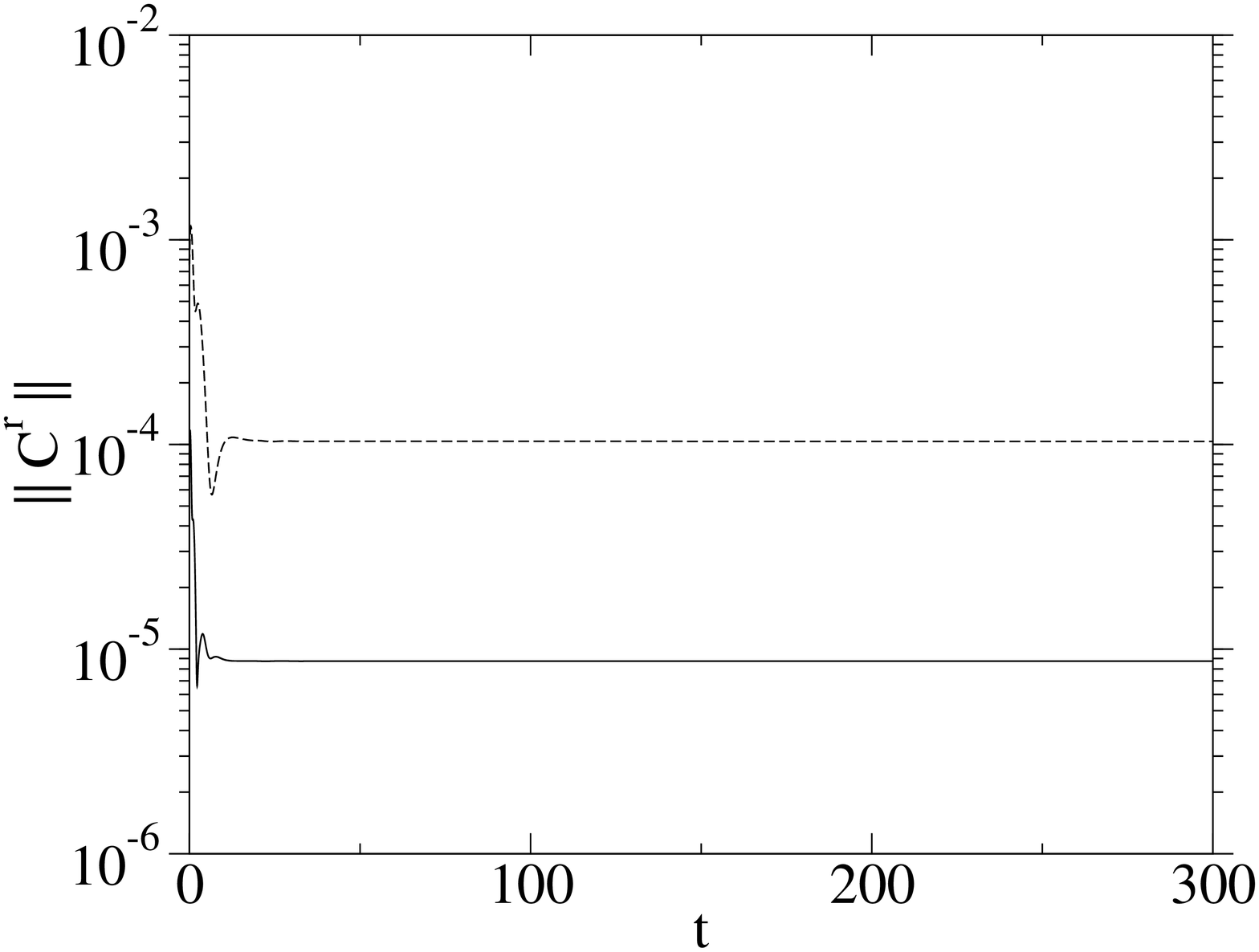}
\includegraphics[scale=0.24]{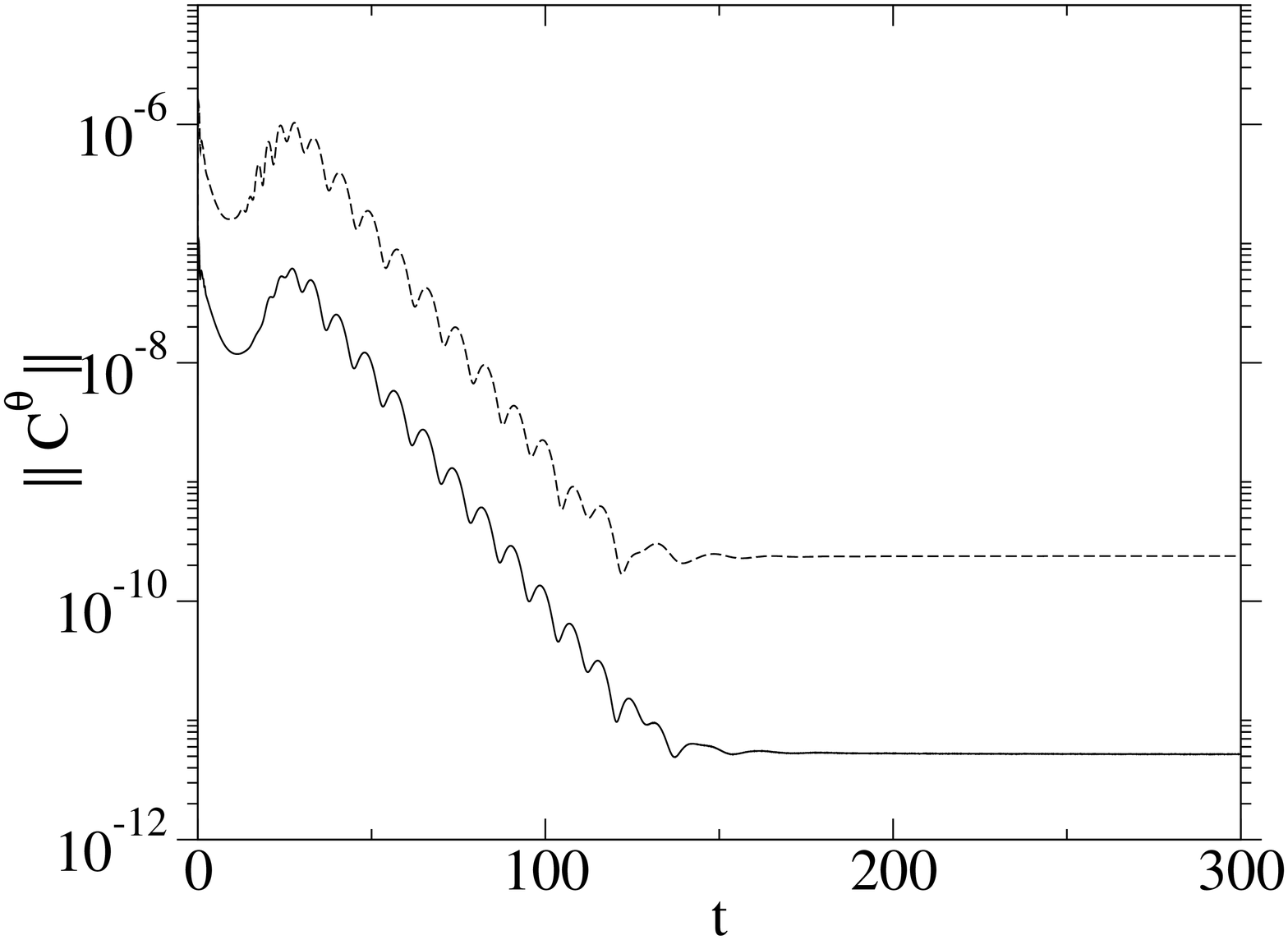} 
\caption{\label{f:Momcons} 
  Momentum constraints \eref{e:Momconsr}--\eref{e:Momconstheta} 
  as functions of time for two different resolutions 
  $(N_r,N_\theta) = (64,8)$ (dashed lines) and $(128,16)$ (solid lines).
  The $r$-momentum constraint (left panel) has been normalized by the
  size of a typical term in \eref{e:Momconsr_Schw} as explained in the text.
}
\end{figure}

In order to demonstrate the consistency and convergence of our code,
we monitor the momentum constraints, which are not solved explicitly
during the evolution.
For the unperturbed Schwarzschild solution, each term in the
$\theta$-momentum constraint \eref{e:Momconstheta} vanishes separately.
This is not true for the $r$-momentum constraint \eref{e:Momconsr};
it vanishes only because of a subtle cancellation of the terms
\begin{equation}
  \label{e:Momconsr_Schw}
  \tfrac{2}{3} \tU_{,r} + \tU(-\tfrac{4}{3} P_r + 2 r^{-1}) = 0
\end{equation}
for the Schwarzschild solution.
Each separate term in this equation is large (in particular close to
the inner boundary) 
and the violation of the $r$-momentum constraint in the perturbed 
evolutions is dominated by the finite-differencing error of these
terms for the Schwarzschild background.
Therefore we have divided the $r$-momentum constraint by
the norm of a typical term in \eref{e:Momconsr_Schw} (roughly $10^2$)
in order to give a better idea of the relative size of the
constraint violation.
The results are shown in figure \ref{f:Momcons}.
The violation of the momentum constraints decreases by just over an
order of magnitude as the resolution is doubled, reasonably close 
to the expected convergence factor of $2^4 = 16$.

Of prime interest is the gravitational radiation emitted by the
perturbed black hole as measured at \Scri.
It can be represented by the Bondi news function \cite{Bondi1962}.
This function was defined originally by studying the asymptotic 
behaviour of the metric in a special coordinate system. 
This coordinate system will in general not agree with the coordinates 
used in a numerical evolution. 
Fortunately, Stewart \cite{Stewart1989} has developed a method for
computing the news in an arbitrary coordinate system,
working as we do here in terms of a conformally related metric.
We begin by specifying a Newman-Penrose \cite{Newman1962}
null tetrad $(l,n,m,\bar m)$ 
that is adapted to \Scri in the sense that
\begin{equation}
  \label{e:AdaptedTetrad}
  m^\alpha \partial_\alpha \psi = 0.
\end{equation}
For definiteness we choose
\begin{equation}
  m = \frac{1}{\sqrt{2} r}\left(\eminus \frac{\partial}{\partial
    \theta} + \rmi \csc\theta \frac{\partial}{\partial\phi}\right).
\end{equation}
Apart from the requirement \eref{e:AdaptedTetrad} the tetrad is arbitrary. 
The resulting expressions will be invariant under spin and boost
transformations.
From the conformal spacetime metric $\four{\gamma}_{\alpha\beta}$ specified in
\eref{e:Conformal4metric} and \eref{e:Conformal3metric}, we compute the 
conformal Ricci tensor $\four{\tilde R}_{\alpha\beta}$. 
The time derivatives in this expression are substituted by using the
evolution equations \eref{e:dtpsi}--\eref{e:dtU}.
The Bondi news function is now given by 
\begin{equation}
  N = \bar\Phi_{02} \equiv \bar m^\alpha \bar m^\beta \four{\tilde R}_{\alpha\beta}.
\end{equation}
The resulting expression is not yet regular at \Scri but it can be
written in a regular form by using the regularity conditions
\eref{e:RegularityConditions}, the boundary condition \eref{e:LapseBC}
and expressions \eref{e:psir} and \eref{e:psirr} for the
radial derivatives of the conformal factor.
Ultimately we obtain
\begin{equation}
  \label{e:news} \fl N \hateq 
  -\eminuss \left[ s(\eta_{,rr} + \eta_{,\theta\theta}) - s^2(\eta_{,r})^2 +
  c\eta_{,\theta} - s^{-1}\eta \right] 
  -\eminus\talpha^{-1}\beta^\theta c\eta_{,r} - \eminus s \tW_{,r}.
\end{equation}

\begin{figure}
\smallskip
\centerline{\includegraphics[scale=0.24]{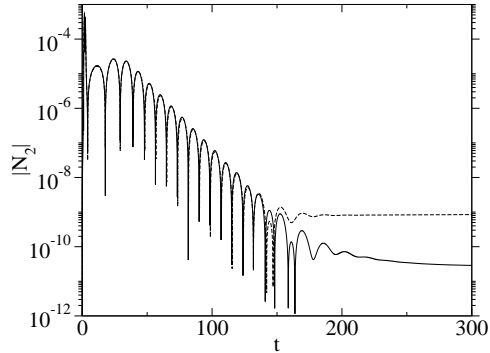}} 
\caption{\label{f:News} 
  $\ell=2$ contribution to the Bondi news function \eref{e:news}, 
  evaluated at \Scri, as a function of time. 
  Numerical results for two different resolutions are shown, 
  $(N_r,N_\theta) = (64,8)$ (dashed line) and $(128,16)$ (solid line).
}
\end{figure}

The Bondi news function has spin weight $-2$ and hence we may expand it
in spin-weighted spherical harmonics \cite{Newman1966} 
${}_{-2} Y_{\ell m}$ with $m=0$ because of the axisymmetry,
\begin{equation}
  N(t,\theta)\vert_{\mathrsfs{I}^+}  
  = \sum_{\ell=2}^\infty N_\ell(t) \; {}_{-2} Y_{\ell 0}(\theta). 
\end{equation}
The $\ell=2$ contribution is plotted in figure \ref{f:News}.
The higher-$\ell$ contributions are considerably smaller (by at least
a factor $10^{-4}$) and have not yet converged for the
numerical resolutions used here.

After an initial peak when the outgoing part of the wave reaches \Scri,
we expect the news to consist of a ringing phase due to the
quasi-normal mode excitations of the black hole, followed by a
polynomial decay (``tail'') due to the backscatter of the radiation off the
curved background spacetime.
These features are apparent in figure \ref{f:News}.
However, whereas the ringing phase has converged quite well, the tail
is still highly resolution-dependent. 
Considerably higher resolutions and/or a higher order of accuracy of
the discretization will be needed in order to study the tail
behaviour.

We expect the quasi-normal mode ringing to have the form
\begin{equation} 
  \label{e:QNMringing} 
  N_\ell \propto \rme^{-\kappa_\ell t} \sin(\omega_\ell t + \varphi_\ell).
\end{equation}
The decay rate $\kappa_\ell$ and frequency $\omega_\ell$ are estimated by
performing a $\chi^2$ fit of \eref{e:QNMringing} 
to the numerical data for the news at \Scri in the time interval 
$60\leqslant t\leqslant 120$.
We find $\kappa_2 = 0.0893 \pm 0.0009$ and $\omega_2 = 0.3738 \pm 0.0012$.
These values correspond to the high-resolution run and the errors have
been estimated by comparing with the low-resolution run.
This numerical result is consistent with the linearized-theory 
result from Leaver's continued-fraction method \cite{Leaver1985}, 
$\kappa_2 = 0.08896$ and $\omega_2 = 0.37367$.


\section{Conclusions}
\label{s:Concl}

The objective of this paper was to demonstrate that stable numerical
evolutions of the Einstein equations on hyperboloidal slices extending
to future null infinity \Scri can be obtained.
While there has been extensive numerical 
work \cite{FrauendienerLRR,Husa2002,Husa2003} using the regular
conformal field equations \cite{Friedrich1983a} on such foliations, 
this is, as far as the author is aware, the first result using 
directly the Einstein equations in a dynamical situation.
The advantage of this approach is that it is more similar to 
formulations of the Einstein equations on (truncated) Cauchy hypersurfaces
that have been used successfully in numerical relativity for a long time.

In this paper, we assumed spacetime to be axisymmetric in order to be
able to experiment with different ways of treating
\Scri numerically in a relatively short computational time.
The axisymmetric reduction of the Einstein equations we used 
(section \ref{s:Formulation}) is
similar to previous works \cite{Rinne2008a,Garfinkle2001,Choptuik2003}
but with a constant-mean-curvature slicing condition instead of
maximal slicing.

The main new difficulty in adapting a formulation of the Einstein equations 
on Cauchy slices to hyperboloidal slices is the treatment of the
formally singular terms in the evolution equations at \Scri.
We applied two different methods \cite{Moncrief2009} in order to obtain regular
versions of these equations at \Scri (section \ref{s:Regularity}).
For the evolutions presented in this paper, we used Penrose's
\cite{Penrose1965} result on the vanishing of the conformal Weyl
tensor at \Scri. 
For this to hold, certain smoothness assumptions on the conformal
metric at \Scri must be satisfied.
An alternate method that does not require these assumptions is based
on Taylor-expanding the fields near \Scri and evaluating the singular
elliptic equations order by order; with this information, the
apparently singular terms in the evolution equations can be evaluated
(\ref{s:TaylorReg}).
This alternate form of the evolution equations at \Scri was found to 
be stable as well and to yield comparable results to within numerical error.
Future work should focus on this latter approach as it requires less
restrictive smoothness assumptions and is most likely compatible
with the so-called polylogarithmic behaviour at \Scri \cite{Andersson1992}.

Our numerical implementation (section \ref{s:NumericalMethod})
uses fairly standard techniques:
fourth-order finite differences, the method of lines with
a fourth-order Runge-Kutta scheme for the evolution equations, and a
multigrid elliptic solver.
The only non-standard modification to the multigrid solver is the use of a
line relaxation in the radial direction in order to deal with the
singular elliptic equations.
With this method, no problems with the convergence of multigrid at the
singular boundary were encountered.
For the evolution equations, the usual Kreiss-Oliger 
dissipation \cite{Kreiss1973} sufficed to ensure stability; 
no special smoothing operations at \Scri were required.

We were able to evolve Schwarzschild spacetime for $1000 M$ ($M$ being
the black hole mass) without any signs of instabilities and with
nearly perfect fourth-order convergence (section \ref{s:NumericalResults}).
Next we included a small gravitational perturbation.
In order to demonstrate the consistency and convergence of the scheme
we monitored the momentum constraints during the evolution.
We derived a regular expression for the Bondi news 
function at \Scri \cite{Bondi1962,Stewart1989} representing the
outgoing gravitational radiation.
From this we computed the frequency and decay rate of the quasi-normal mode
radiation and found good agreement with the tabulated values
from linearized theory.
The numerical resolutions we were able to reach were not yet high
enough in order to study the subsequent polynomial decay (the ``tail'').

Currently the code is too slow in order to make runs with higher
resolutions feasible (the run at the higher resolution in section 
\ref{s:Schw} took about two weeks on a single processor).
Not much effort has been spent on optimizing the computational
efficiency of the code and there is certainly some room for
improvement here.
It seems likely, however, that a finite-difference method of higher
than fourth order or a pseudospectral method will need to be used in
order to reach the computational accuracy needed to study tails.

Although our multigrid solver has optimal computational complexity 
(linear in the number of unknowns) \cite{Brandt1977,BriggsMG},
the fact that our formulation contains elliptic equations is a
major drawback on the computational speed.
This is likely to be even more severe in the case without spacetime
symmetries.
From this point of view, a hyperbolic formulation such as the one by
Zengino\u{g}lu \cite{Zenginoglu2008} seems advantageous.
It would be interesting to see whether the methods of
\cite{Moncrief2009} can be applied to that formulation as well so that
regular versions of the evolution equations at \Scri are obtained.

In this first study we imposed axisymmetry in order to be able to
experiment with different numerical treatments of \Scri more quickly.
The next step will be the extension to spacetimes without any symmetries.
As far as the numerical stability of \Scri is concerned, we do not
expect any new difficulties here because unlike spherical symmetry,
axisymmetry already shares with the non-symmetric case 
what we believe are the key challenges, namely the existence of a
dimension tangential to the conformal boundary and the presence of
gravitational radiation.
It remains to be seen which formulation of the Einstein equations will
be more successful for this purpose.
We expect the ideas and methods developed in this work to
be applicable to a variety of formulations and gauge choices.


\ackn
I am grateful to Vincent Moncrief for numerous fruitful discussions.
Helpful interactions with James Bardeen, Luisa Buchman, Harald Pfeiffer, 
Olivier Sarbach, John Stewart and An{\i}l Zengino\u{g}lu are much appreciated.
I am supported by a Research Fellowship at King's College Cambridge.


\appendix

\section{Further details on the solution of the elliptic equations}
\label{s:EllipticDetails}


\subsection{Quasi-isotropic gauge conditions}
\label{s:Shift}

The quasi-isotropic gauge conditions
\eref{e:Shift1r} and \eref{e:Shift1theta} are formally equivalent to
the (inhomogeneous) Cauchy-Riemann equations.
In order to solve them, one usually combines derivatives of these
equations to obtain two decoupled Poisson equations, which can easily
be solved.
However, care must be taken so that the solution to the second-order Poisson
equations is in fact also a solution to the original first-order 
Cauchy-Riemann equations (in general the two may differ by integration 
constants).
We use the following approach.

The quasi-isotropic gauge conditions \eref{e:Shift1r} and 
\eref{e:Shift1theta} are
\begin{eqnarray}
  \label{e:Sr}
  0 = \cS^r \equiv \beta^r{}_{,r} - \beta^\theta{}_{,\theta} - r^{-1}
  \beta^r - \talpha\tU,\\
  \label{e:Stheta}
  0 = \cS^\theta \equiv \beta^r{}_{,\theta} + r^2
  \beta^\theta{}_{,r} - 2 \talpha\tKrth.
\end{eqnarray}
Derivatives of these are combined to obtain a scalar Poisson equation,
\begin{eqnarray}
  \label{e:Shift2theta}
  \fl 0 = -\cS^r{}_{,\theta} + r^{-2} \cS^\theta{}_{,r} = &
  \beta^\theta{}_{,rr} + r^{-1} \beta^\theta{}_{,r}
  + r^{-2} \beta^\theta{}_{,\theta\theta}
  - 2 r^{-1} \talpha_{,r} \tKrth + r^{-2} \talpha_{,\theta} \tU
  \nonumber\\
  &- 2 r^{-2} \talpha {\tKrth}_{,r} + r^{-2} \talpha \tU_{,\theta} + 2
  r^{-3} \talpha \tKrth.
\end{eqnarray}
This equation is solved for $\beta^\theta$ using multigrid, with
homogeneous Dirichlet boundary conditions.

However, we do \emph{not} solve the corresponding Poisson equation 
for $\beta^r$.
Instead, we observe that once $\beta^\theta$ is known we can
immediately integrate \eref{e:Sr} and \eref{e:Stheta} to find $\beta^r$.
First, \eref{e:Stheta} is integrated along the outer boundary \Scri
to determine $\beta^r$ there. 
The integration constant can be chosen arbitrarily, 
e.g.~by requiring the average of $\beta^r$ along \Scri to agree 
with its value for the exact Schwarzschild solution.
Next, \eref{e:Sr} is integrated from each point on the outer boundary
towards the interior.
In this way we ensure that $\cS^r = 0$ everywhere. 
We have also enforced $\cS^\theta \hateq 0$ at the boundary.
What remains to be shown is that $\cS^\theta = 0$ in the interior.
But this follows immediately from \eref{e:Shift2theta}:
given that we already have $\cS^r = 0$ everywhere, this equation
reduces to $\cS^\theta{}_{,r} = 0$, and together with the initial
condition $\cS^\theta \hateq 0$ this implies $\cS^\theta = 0$ everywhere.

Apart from ensuring that the first-order conditions
\eref{e:Sr} and \eref{e:Stheta} are indeed satisfied, we also avoid
solving another elliptic equation (for $\beta^r$) and thus save
computational time.


\subsection{Momentum constraints}
\label{s:Momcons}

For the intial data only, the momentum constraints are solved using a
method similar to the one due to York \cite{York1979}.
We introduce a vector $V^A$ and write the extrinsic curvature
variables as
\begin{eqnarray}
  \label{e:Usub}
  \tU = \psi^2 (V^r{}_{,r} - V^\theta{}_{,\theta} - r^{-1} V^r) \equiv
  \psi^2 V_-,\\
  \label{e:Krthsub}
  \tKrth = \half \psi^2 (V^r{}_{,\theta} + r^2 V^\theta{}_{,r}) \equiv
  \half \psi^2 V_+,\\ 
  \label{e:Wsub}
  \tW = \psi^2 \hW.
\end{eqnarray}
Note the similarity of \eref{e:Usub} and \eref{e:Krthsub} with
\eref{e:Shift1r} and \eref{e:Shift1theta}. 
The field $\hW$ can be freely specified.
The substitutions \eref{e:Usub}--\eref{e:Wsub} completely decouple the
momentum constraints \eref{e:Momconsr} and \eref{e:Momconstheta} 
from the remaining elliptic equations:
\begin{eqnarray}
  \fl 0 = \tfrac{2}{3} V^r{}_{,rr} + \half r^{-2} V^r{}_{,\theta\theta} -
  \tfrac{1}{6} V^\theta{}_{,r\theta} - \tfrac{2}{3}
  V^\theta{}_{,\theta} + \third s \hW_{,r} + r^{-1} s \hW \nonumber \\
  + V_- (s\eta_{,r} + \tfrac{4}{3} r^{-1}) + r^{-2} V_+
  (s\eta_{,\theta} + c \eta + \half \cs),\\
  \fl 0 = \tfrac{1}{6} V^r{}_{,r\theta} + \half r^2 V^\theta{}_{,rr} +
  \third V^\theta{}_{,\theta\theta} - \tfrac{2}{3} r^{-1} V^r{}_{,\theta}
  + \third s \hW_{,\theta} + \tfrac{4}{3} c \hW \nonumber\\
  - V_- (s\eta_{,\theta} + c \eta) + V_+ (s\eta_{,r} + 2 r^{-1}).
\end{eqnarray}

We take $V^A$ to agree with the corresponding vector computed from
the exact Schwarzschild solution at the radial boundaries, in particular
we have $V^A \hateq 0$ at \Scri, and $V^\theta$ also vanishes at the
inner boundary.
The angular parity of $V^A$ is identical with that of $\beta^A$
(table \ref{t:Parities}).


\section{Alternate regularized evolution equations}
\label{s:TaylorReg}

In this appendix we derive alternate regular forms of the evolution
equations at \Scri using the Taylor expansion method developed in
\cite{Moncrief2009}.
Within a hyperboloidal hypersurface $t=\const$, all the fields are 
expanded in finite Taylor series about \Scri ($r=1$ in our case),
\begin{equation}
  \label{e:Taylor}
  u(x^i) = \sum_{n=0}^N \frac{1}{n!} u_n(\theta,\phi) (r-1)^n, \qquad
  u_n = \lim_{r \nearrow 1} \partial_r^n u.
\end{equation}
These series are substituted in the singular elliptic equations and
evaluated order by order.
Again we have found a computer algebra programme \cite{ReduceManual} 
to be very helpful here.
From the Hamiltonian constraint \eref{e:Hamcons} we obtain
\begin{eqnarray}
  \label{e:psir}
  \fl \psi_{,r} \hateq -\third K \eplus,\\
  \label{e:psirr}
  \fl \psi_{,rr} \hateq -K \eplus (\half s\eta_{,r} + \third),\\
  \fl \psi_{,rrr} \hateq K \eplus \left\{ \tfrac{2c}{3s}
    (s\eta)_{,\theta} + \third (s\eta)_{,\theta\theta} - \third \left[
      (s\eta)_{,\theta}\right]^2 + \tfrac{2}{3} s \eta_{,rr} +
    \tfrac{5}{12} (s \eta_{,r})^2 + \tfrac{2}{3} s \eta_{,r} \right\}.
\end{eqnarray}
The slicing condition \eref{e:CMClapse} yields
\begin{eqnarray}
  \fl \talpha_{,r} \hateq \talpha (\half s \eta_{,r} + 1),\\
  \fl \alpha_{,rr} \hateq \talpha \left[ \tfrac{3c}{2s}
    (s\eta)_{,\theta} + \half (s\eta)_{,\theta\theta} + \half s
    \eta_{,rr} - \half (s\eta_{,r})^2 + s\eta_{,r} + 1\right]\nonumber\\
  + \tfrac{c}{2s} \talpha_{,\theta} + \half \talpha_{,\theta\theta}
  - \tfrac{3}{2} \talpha_{,\theta} (s\eta)_{,\theta}.
\end{eqnarray}
From the momentum constraints \eref{e:Momconsr} and
\eref{e:Momconstheta} we obtain, respectively,
\begin{eqnarray}
  \label{e:Wr}
  s \tW_{,r} \hateq -2\tU_{,r} + \tfrac{3}{2} \eminus (s\eta_{,r})^2,\\
  \label{e:Krthr}
  \tKrth_{,r} \hateq \eminus \left[ -c\eta_{,r} - \half
    (s\eta_{,r})_{,\theta} + s \eta_{,r} (s\eta)_{,\theta} \right].
\end{eqnarray}

Substituting the Taylor expansions \eref{e:Taylor} 
in the right-hand sides of the 
evolution equations \eref{e:dtW}--\eref{e:dtU} and using the above
results \eref{e:psir}--\eref{e:Krthr}
for the radial derivatives of the fields, together with the
regularity conditions \eref{e:RegularityConditions} and the boundary
condition \eref{e:LapseBC}, we obtain the following regular forms of
the evolution equations at \Scri.
\begin{eqnarray}
  \label{e:dtWreg2}
  \fl \tW_{,t} \hateq \eminuss \big\{ \talpha \left[ c (s^{-1}\eta)_{,\theta}
      + \eta_{,\theta\theta} + \eta_{,rr} - s (\eta_{,r})^2 - \eta \right]
    + \talpha_{,\theta}(\tfrac{c}{s}\eta + \eta_{,\theta})\nonumber\\ 
    - (s^{-1} \talpha_{,\theta})_{,\theta} \big\}
  + \eminus\talpha\tW_{,r} 
  + \eminus\beta^\theta[\eta_{,r} 
    (c \eta - 2\tfrac{c}{s} + s \eta_{,\theta}) - \eta_{,r\theta}] ,\\
  \label{e:dtKrthreg2}
  \fl \tKrth_{,t} \hateq 0,\\
  \label{e:dtUreg2}
  \fl \tU_{,t} \hateq -\half s\tW_{,t}.
\end{eqnarray}
Notice in particular how \eref{e:dtKrthreg2} and \eref{e:dtUreg2} are
compatible with the regularity conditions \eref{e:RegularityConditions}.

\begin{figure}
\smallskip
\centerline{\includegraphics[scale=0.24]{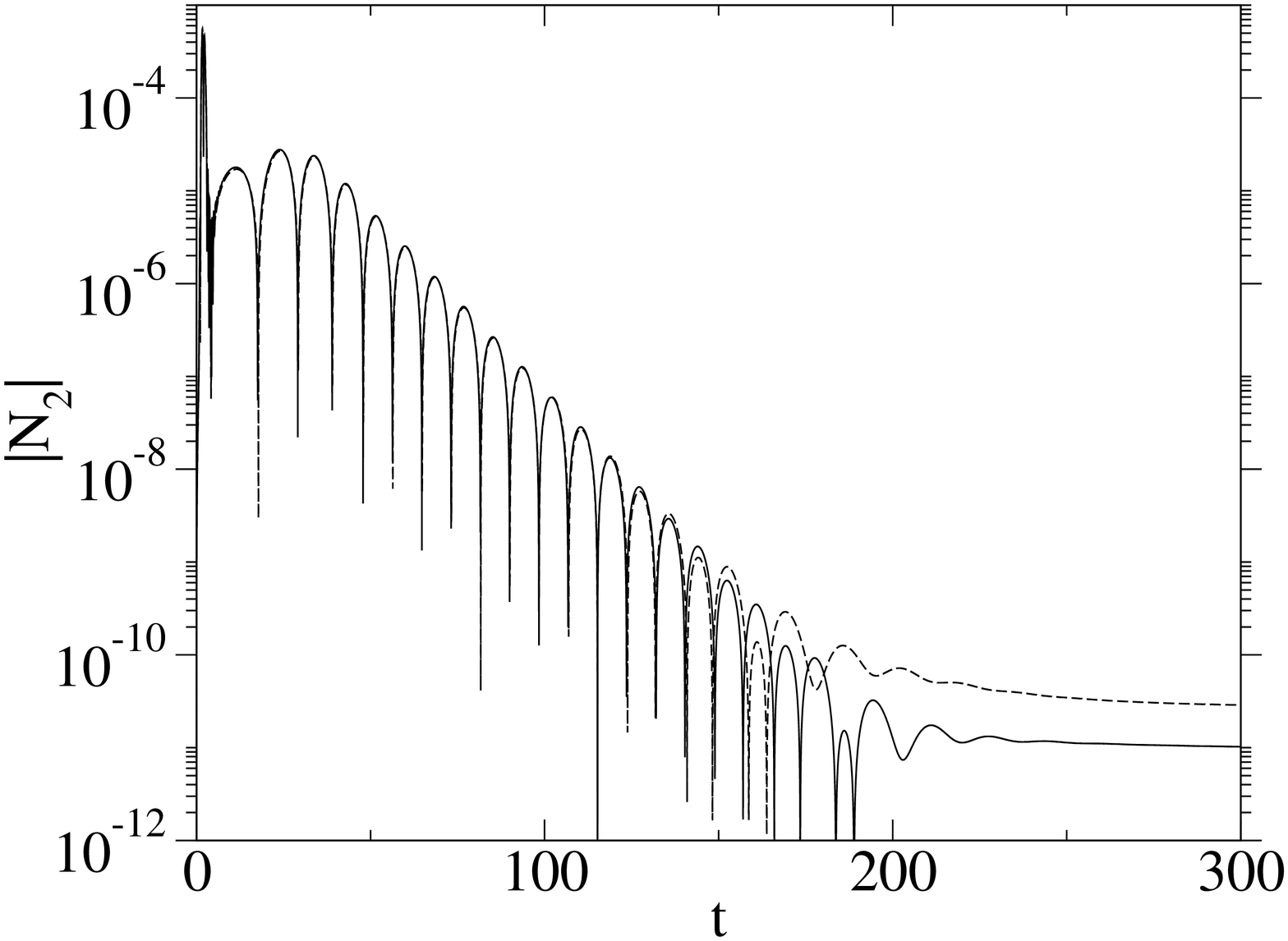}} 
\caption{\label{f:NewsTaylorReg} 
  $\ell=2$ contribution to the Bondi news function \eref{e:news}
  at \Scri for an evolution using regularization at \Scri based on
  Taylor expansions (solid line) as compared with the approach based
  on Penrose's smoothness assumption (dashed line) that was used 
  in section \ref{s:NumericalResults}.
  The numerical resolution is $(N_r,N_\theta) = (128,16)$.
}
\end{figure}

Figure \ref{f:NewsTaylorReg} shows the Bondi news function at \Scri 
for a numerical evolution using the above Taylor-regularized 
evolution equations at \Scri.
The result agrees well with the evolution based on the Penrose
regularization technique that was discussed in section \ref{s:Regularity}
and used for the evolution shown in figure \ref{f:News}.
The discrepancy for times $t \gtrsim 150$ can be attributed to
numerical error---figure \ref{f:News} shows that the solution is
still highly resolution-dependent at those late times.
The corresponding curves in figures \ref{f:Schw} and \ref{f:Momcons}
virtually overlap for the two regularization techniques.

\section{Finite-difference operators}
\label{s:FD}

In order to compute derivatives with respect to $r$, we first compute
finite-difference approximations to derivatives with respect to the
map coordinate $x$ and then apply the chain rule,
\begin{equation}
  u_{,r} = \frac{\rmd x}{\rmd r} u_{,x} , \qquad
  u_{,rr} = \left(\frac{\rmd x}{\rmd r}\right)^2 u_{,xx} 
  + \frac{\rmd^2 x}{\rmd r^2} \, u_{,x}.
\end{equation}
Here $\rmd x/ \rmd r$ and $\rmd^2 x/ \rmd r^2$ are computed
analytically from the given form of the mapping $r(x)$, equation 
\eref{e:RadialMap}.

Fourth-order accurate finite difference operators are used with
respect to both $x$ and $\theta$.
It suffices to display their one-dimensional forms here.
The centred finite difference operators are
\begin{eqnarray}
  u'_i \rightarrow\tfrac{1}{12 h} 
  (u_{i-2} - 8 u_{i-1} + 8 u_{i+1} - u_{i+2}),\\
  u''_i \rightarrow\tfrac{1}{12 h^2} 
  (-u_{i-2} + 16 u_{i-1} - 30 u_i + 16 u_{i+1} - u_{i+2}),
\end{eqnarray}
where $h$ denotes the (uniform) grid spacing and the indices 
refer to the grid points where the quantities are evaluated.
The forward differences are
\begin{eqnarray}
  u'_0 \rightarrow \tfrac{1}{12 h} 
  (-25 u_0 + 48 u_1 - 36 u_2 + 16 u_3 - 3 u_4),\\
  u'_1 \rightarrow \tfrac{1}{12 h} 
  (-3 u_0 - 10 u_1 + 18 u_2 - 6 u_3 + u_4),\\
  u''_0 \rightarrow \tfrac{1}{12 h^2}
  (45 u_0 - 154 u_1 + 214 u_2 - 156 u_3 + 61 u_4 - 10 u_5),\\
  u''_1 \rightarrow \tfrac{1}{12 h^2}
  (10 u_0 - 15 u_1 - 4 u_2 + 14 u_3 - 6 u_4 + u_5),
\end{eqnarray}
and expressions for the backward differences follow by symmetry.

In the $x$-direction we use centred differences for $2\leqslant i
\leqslant N_r-2$, forward differences at $i=0,1$ and backward
differences at $i=N_r,N_r-1$.
In the $\theta$-direction centred differences are used at all
interior grid points $1\leqslant j \leqslant N_\theta$
(note that ghost points are provided at $j=-1,0$ and 
$j=N_\theta+1, N_\theta+2$).

The expression $(s^{-1} u)_{,\theta}$ is discretized by
finite-differencing the function $s^{-1} u$ with respect to $\theta$.
No such special differencing is used for 
$(s^{-1} u_{,\theta})_{,\theta}$, which we simply expand into
$s^{-1} u_{,\theta\theta} - s^{-2} c u_{,\theta}$ and then
discretize in the standard way.
Mixed derivatives $u_{,x\theta}$ are discretized by applying the
finite-difference operators with respect to $x$ and $\theta$ 
subsequently in the obvious way.

We use Kreiss-Oliger dissipation \cite{Kreiss1973}, whereby the operator
\begin{eqnarray}
  \fl (Q u)_i = \left(\tfrac{\epsilon}{64} h^5 D_+^3 D_-^3 u\right)_i 
  \nonumber \\ 
  = \tfrac{\epsilon}{64 h} (u_{i-3} - 6 u_{i-2} + 15 u_{i-1} - 20 u_i +
  15 u_{i+1} - 6 u_{i+2} + u_{i+3}) 
\end{eqnarray}
is applied to each variable $u$ both in the $x$ and the $\theta$
direction and the results added to the discretized right-hand side 
of the evolution equation for $u$.
Here $D_\pm$ denote the first-order forward and backward 
finite-differencing operators, and a typical choice of the dissipation 
parameter is $\epsilon=1/2$.
In the $x$-direction the above dissipation operator is added at grid
points $2 \leqslant i \leqslant N_r-2$ and in the $\theta$-direction
at grid points $2 \leqslant j \leqslant N_\theta-1$.
Including $j=2,N_\theta-1$ has been found to be essential in order to
eliminate an angular instability. 


\section*{References}

\providecommand{\newblock}{}

\end{document}